\documentclass[pss,fleqn,embeddedheads]{w-art}
\usepackage{times}
\usepackage{w-thm}
\usepackage{amssymb,cite}
\usepackage[]{graphicx}
\newcommand{\figwidth}{0.95\columnwidth}
\begin{document}
\DOIsuffix{theDOIsuffix}
\Volume{XX}
\Issue{1}
\Month{01}
\Year{2003}
\pagespan{3}{}
\Receiveddate{21 August 2005}
\Reviseddate{27 August 2005}
\Accepteddate{}
\Dateposted{$Revision: 1.12 $, compiled \today}



\title[Sequence Dependence in DNA]{Sequence Dependence of Electronic Transport in DNA}


\author[Antonio Rodriguez]{Antonio Rodriguez\inst{1}}
\address[\inst{1}]{Dpto Matem\'{a}tica Aplicada y Estabística, E.U.I.T.
  Aeron\'{a}utics. U.P.M., PZA Cardenal Cisneros s/n, Madrid 28040,
  Spain}
\author[Rudolf A.\ R\"{o}mer]{Rudolf A.\ R\"{o}mer\footnote{Corresponding
     author: e-mail: {\sf r.roemer@warwick.ac.uk}, Phone: +44\,2476\,574\,328,
     Fax: +44\,7876\,858\,246}\inst{2}}
\address[\inst{2}]{Physics Department \& Centre for Scientific
Computing, University of Warwick, Coventry CV4 7AL, U.K.}
\author[Matthew S.\ Turner]{Matthew S.\ Turner\inst{2}}
\begin{abstract}
  We study electronic transport in long DNA chains using the
  tight-binding approach for a ladder-like model of DNA. We find
  insulating behavior with localizaton lengths $\xi \approx 25$ in units
  of average base-pair seperation.  Furthermore, we observe small, but
  significant differences between $\lambda$-DNA, centromeric DNA,
  promoter sequences as well as random-ATGC DNA.
\end{abstract}
\maketitle                   





\section{Introduction}
\label{sec-introduction}

DNA is a macro-molecule consisting of repeated stacks of bases
formed by either AT (TA) or GC (CG) pairs coupled via hydrogen
bonds and held in the double-helix structure by a sugar-phosphate
backbone. In most models of electronic transport
\cite{CunCPD02,Zho03} it has been assumed --- following earlier
pioneering work \cite{LadSOB86,BakOLS86} --- that the transmission
channels are along the long axis of the DNA molecule and that the
conduction path is due to $\pi$-orbital overlap between
consecutive bases \cite{TreHB01}.

A simple quasi-1D model incorporating these aspects has been recently
introduced \cite{KloRT05}, building on an earlier, even simpler 1D model
\cite{CunCPD02}. For the model, electronic transport properties have
been investigated in terms of localisation lengths \cite{Yam04, KloRT05},
crudely speaking the length over which electrons travel.  Various types
of disorder, including random potentials, had been employed to account
for different real environments. It was found that random and
$\lambda$-DNA have localisation lengths allowing for electron motion
among a few dozen base pairs only. However, poly(dG)-poly(dC) and also
telomeric-DNA have much larger electron localization lengths. In Ref.\ \cite{KloRT05},
a novel enhancement of localisation lengths has been observed at
particular energies for an {\em increasing} binary backbone disorder.

\section{The DNA tight-binding model}
\label{sec-TB-model}

A convenient tight binding model for DNA can be constructed as
follows: it has two central conduction channels in which
individual sites represent an individual base; these are
interconnected and further linked to upper and lower sites,
representing the backbone, but are \emph{not} interconnected along
the backbone. Every link between sites implies the presence of a
hopping amplitude. The Hamiltonian $H_L$ for this ladder-like
model is given by
\begin{eqnarray}
H_{L} &=& \sum_{i=1}^{L}
 \sum_{\tau=1,2}
    \left( t_{i,\tau}|i,\tau\rangle \langle i+1,\tau| +
    \varepsilon_{i,\tau} |i,\tau\rangle \langle i,\tau| \right)
+ \sum_{q=\uparrow,\downarrow}
    \left( t_i^q |i,\tau\rangle \langle i,q(\tau)|+
    \varepsilon_i^q|i,q\rangle \langle i,q| \right)
\nonumber \\
 & & + \sum_{i=1}^{L}
 t_{1,2}|i,1\rangle \langle i,2|
  \label{eq-ham2D}\label{eq-ladder}
\end{eqnarray}
where $t_{i,\tau}$ is the hopping amplitude between sites along
each branch $\tau=1$, $2$ and $\varepsilon_{i,\tau}$ is the
corresponding onsite potential energy. $t_i^q$ and and
$\varepsilon_i^q$ give hopping amplitudes and onsite energies at
the backbone sites. Also, $q(\tau)=\uparrow, \downarrow$ for
$\tau=1, 2$, respectively. The parameter $t_{12}$ represents the
hopping between the two central branches, i.e., perpendicular to
the direction of conduction. Quantum chemical calculations with
semi-empirical wave function bases using the SPARTAN package
\cite{Spartan} results suggest that this value, dominated by the
wave function overlap across the hydrogen bonds, is weak and so we
choose $t_{12}= 1/10$. As we restrict our attention here to pure
DNA, we also set $\varepsilon_{i,\tau}=0$ for all $i$ and $\tau$.

The model (\ref{eq-ladder}) clearly represents a dramatic
simplification of DNA. Nevertheless, in Ref.\ \cite{CunCPD02} it
had been shown that an even simpler model --- in which base-pairs
are combined into a single site --- when applied to an artificial
sequence of repeated GC base pairs, poly(dG)-poly(dC) DNA,
reproduces experimental data current-voltage measurements when
$t_{i}=0.37 e$V and $t_i^q=0.74 e$V are being used. This motivates
the above parametrization of $t_i^q = 2 t_{i}$ and
$t_{i,\tau}\equiv 1$ for hopping between like (GC/GC, AT/AT)
pairs. Assuming that the wave function overlap between consecutive
bases along the DNA strand is weaker between unlike and
non-matching bases (AT/GC, TA/GC, etc.) we thus choose $1/2$.
Furthermore, since the energetic differences in the adiabatic
electron affinities of the bases are small \cite{WesLPS01}, we
choose $\varepsilon_{i}=0$ for all $i$. Due to the
non-connectedness of the backbone sites along the DNA strands, the
model (\ref{eq-ladder}) can be further simplified to yield a model
in which the backbone sites are incorporated into the electronic
structure of the DNA. The effective ladder model reads as
\begin{equation}
\tilde{H}_{L} = \sum_{i=1}^{L}
  t_{1,2}|i,1\rangle \langle i,2| +
\sum_{\tau=1,2}
    t_{i,\tau}|i,\tau\rangle \langle i+1,\tau|
 + \left[ \varepsilon_{i,\tau} -
 \frac{\left(t_{i}^{q(\tau)}\right)^{2}}{\varepsilon_{i}^{q(\tau)} - E}
\right] |i,\tau\rangle \langle i,\tau| + h.c. \quad .
\label{eq-ladder-effective}
\end{equation}
Thus the backbone has been incorporated into an {\em
energy-dependent} onsite potential on the main DNA sites. This
effect is at the heart of the enhancement of localization lengths
due to increasing binary backbone disorder reported previously
\cite{KloRT05}.

\section{$\lambda$-DNA, centromers and promoters}
\label{sec-DNA}

We shall use 2 naturally occurring DNA sequences (``strings").  (i)
$\lambda$-DNA \cite{lambda} is DNA from the bacteriophage virus. It has
a sequence of 48502 base pairs and is biologically very well
characterised. Its ratio $\alpha$ of like to un-like base-pairs is
$\alpha_{\lambda}=0.949$. (ii) centromeric DNA for chromosome 2 of yeast
has $813138$ base pairs \cite{cen2} and $\alpha_{{\rm centro.}}=0.955$.
This DNA is also rich in AT bases and has a high rate of repetitions
which should be favourable for electronic transport.

Another class of naturally existing DNA strands is provided by so-called
promoter sequences. We use a collection of 4986 is these which have been
assembled from the TRANSFAC database and cover a range of organisms such
as mouse, human, fly, and various viruses.  Promoter sequences are
biologically very interesting because they represent those places along
a DNA string where polymerase enzymes bind and start the copying process
that eventually leads to synthesis of proteins. On average, these
promoters consist of approximetely $17$ base-pairs, much too short for a
valid localization length analysis by TMM. Therefore, we concatenate
them into a $86827$ base-pair long {\em super-promoter} with
$\alpha_{{\rm super-p.}}=0.921$. In order to obtain representative
results, $100$ such super-promoters have been constructed, representing
different random arrangements of the promoters, and the results
presented later will be averages\footnote{Averages of $\xi$ are computed
  by averaging $1/\xi$.}.

Occasionally, we show results for ``scrambled'' DNA. This is DNA with
the same number of A, T, C, G bases, but with their order randomised.
Clearly, such sequences contain the same set of electronic potentials
and hopping variations, but would perform quite differently in a biological context. A comparison of their transport properties with those
from the original sequence thus allows to measure how important the
exact fidelity of a sequence is. On average, we find for these sequences
$\alpha_{\lambda/{\rm S}}=0.899$, $\alpha_{{\rm centro./S}}= 0.9951$ and
$\alpha_{{\rm super-p./S}}= 0.901$.

A convenient choice of artificial DNA strand is a simple, $100000$
base-pair long {\em random} sequence of the four bases, random-ATGC DNA,
which we construct with equal probability for all 4 bases ($\alpha_{{\rm
    random}}=0.901$). We shall also `promote' these random DNA strings
by inserting all $4086$ promoter sequences at random positions in the
random-ATGC DNA ($\alpha_{{\rm random/P}}= 0.910$).

\section{Results for localization lengths}
\label{sec-results}

For studying the transport properties of model (\ref{eq-ladder}),
we use a variant of the iterative transfer-matrix method (TMM)
\cite{PicS81a,PicS81b,MacK83,KraM93,Mac94}. The TMM allows us to
determine the localisation length $\xi$ of electronic states in
the present system with fixed cross sections $M=2$ (ladder) and
length $L \gg M$. Traditionally, a few million sites are needed
for $L$ to achieve reasonable accuracy for $\xi$. However, in the
present situation we are interested in finding $\xi$ also for much
shorter DNA strands of typically only a few ten thousand base-pair
long sequences.  Thus in order to restore the required precision,
we have modified the conventional TMM and can now perform TMM on a
system of fixed length $L_0$ by repeating forward- and
backward-TMM steps \cite{FraMPW95,RomS97b,NdaRS04,KloRT05}.

We have computed the energy dependence of the localization lengths
for all sequences of section \ref{sec-DNA}. In addition,
$\lambda$-DNA, centromeric DNA and the super-promoter DNA where
also scrambled $100$ times and the localization length of each
resulting sequence measured and the appropriate average
constructed. Also, we constructed $100$ promoted random-ATGC DNA
sequences.
\begin{figure}
  \centering
  \includegraphics[width=\figwidth]{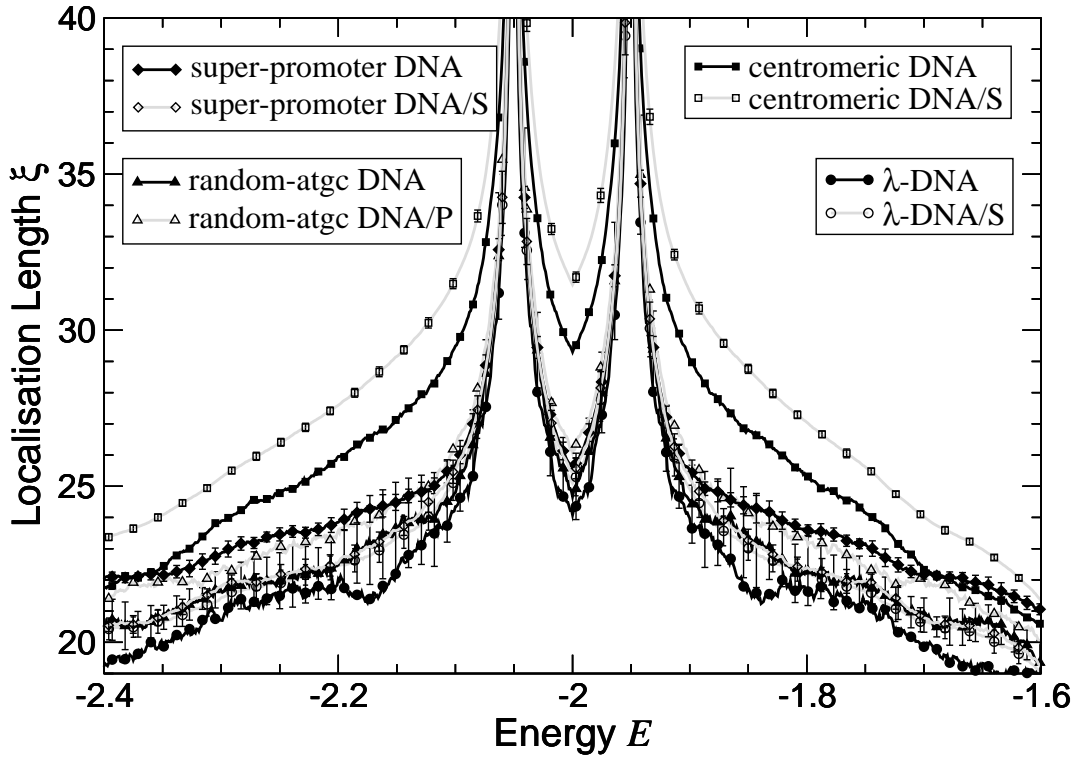}
  \caption{\label{fig-DD-Loc_Energy}
    Localization lengths $\xi$ versus Fermi energy $E$ for various clean
    DNA (solid symbols), scrambled DNA (DNA/S, (open $\diamond$,
    $\Box$, $\circ$) and promoted DNA (DNA/R, open $\triangle$)
    strands. Only every 10th (20th) symbol is shown for clean
    (scrambled/promoted) DNA. Error bars reflect the standard deviation
    after sampling the different sequences for random-ATGC, scrambled
    and promoted DNA.}
\end{figure}
As shown previously \cite{KloRT05}, the energy dependence of $\xi$
reflects the backbone-induced two-band structure. The obtained $\xi(E)$
values for the lower band are shown in Fig.\ \ref{fig-DD-Loc_Energy}. In
the absence of any onsite-disorder, we find two prominent peaks
separated by $t_{1,2}$ and $\xi(E)=\xi(-E)$. We also see that
$\lambda$-DNA has roughly the same $\xi(E)$ dependence as
random-ATGC-DNA. Promoting a given DNA sequence leads to small increases
in localization length $\xi$, whereas scrambling can lead to increase as
well as decrease.  The super-promoter has larger $\xi$ values compared
to random-atcg- and $\lambda$-DNA. Most surprisingly, centromeric DNA
--- the longest investigated DNA sequence --- has a much larger
localization length than all other DNA sequences and this even {\em
  increases} after scrambling.

\section{Conclusions}
\label{sec-conclusions}

We have shown that the ladder model (\ref{eq-ladder}) is a simple, yet
non-trivial representation of DNA within the tight-binding formalism.
While keeping the number of parameters small, we manage to reproduce the
wide-gap structure observed in much more accurate quantum chemical
calculations of short DNA strands
\cite{DavI04,BhaBB03,CunCPD02,GarABG01,RakAPK01}. In order to study the
transport properties, we employ a variant of the TMM which provides
useful information about the spatial extend $\xi$ of electronic states
along a DNA strand in the quantum regime at $T=0$. We note that the
values of $\xi$ which we find are around $25$ in the band which is
surprisingly close to studies of range dependence of electron transfer
\cite{WanFSB00,BooLCD03,KelB99,MurAJG93,OneDB04,DelB03,TreHB01}.

From our results, we find clear differences in localization lengths which are not
simple related to a difference in DNA composition, but also reflect the order of
base-pairs. Still the differences are within $10-20\%$ and it remains unclear how
relevant these findings are biologically, i.e., whether electronic transport
plays a role in the biological mechanism of DNA repair and protein generation.

\begin{acknowledgement}
  This work has been supported in part by the Royal Society. We thank
  A.\ Croy and C.\ Sohrmann for useful discussions.
\end{acknowledgement}


\end{document}